\documentclass[%
 aip,
 amsmath,amssymb,
preprint,%
]{revtex4-2}

\usepackage{graphicx}
\usepackage{dcolumn}
\usepackage{bm}

\usepackage[utf8]{inputenc}
\usepackage[T1]{fontenc}
\usepackage{mathptmx}
\usepackage{xcolor}

\DeclareMathAlphabet{\mathsfit}{\encodingdefault}{\sfdefault}{m}{sl}
\SetMathAlphabet{\mathsfit}{bold}{\encodingdefault}{\sfdefault}{bx}{sl}
\newcommand{\vect}[1]{\bm{#1}}

\newcommand{\apjl}{The Astrophysical Journal Letters}

\newcommand{\aap}{Astronomy \& Astrophysics}
\newcommand{\jgr}{Journal of Geophysical Research}
\newcommand{\grl}{Geophysical Research Letters}
\newcommand{\solphys}{Solar Physics}
\newcommand{\ssr}{Space Science Reviews}
\newcommand{\araa}{Annual Review of Astronomy and Astrophysics}
\newcommand{\planss}{Planetary Space Science}

\begin{document}

\preprint{AIP/123-QED}

\title[Power-law Formation]{The Acceleration of Charged Particles and Formation of Power-law Energy Spectra in Nonrelativistic Magnetic Reconnection}

\author{Xiaocan Li}
\email{Xiaocan.Li@dartmouth.edu.}
\affiliation{Department of Physics and Astronomy, Dartmouth College, Hanover, NH 03755, USA}%
\author{Fan Guo}%
\affiliation{Los Alamos National Laboratory, Los Alamos, NM 87545, USA}%
\author{Yi-Hsin Liu}
\affiliation{Department of Physics and Astronomy, Dartmouth College, Hanover, NH 03755, USA}%


\date{\today}

\begin{abstract}
    Magnetic reconnection is a primary driver of particle acceleration processes in space and astrophysical plasmas. Understanding how particles are accelerated and the resulting particle energy spectra is among the central topics in reconnection studies. We review recent advances in addressing this problem in nonrelativistic reconnection that is relevant to space and solar plasmas and beyond. We focus on particle acceleration mechanisms, particle transport due to 3D reconnection physics, and their roles in forming power-law particle energy spectra. We conclude by pointing out the challenges in studying particle acceleration and transport in a large-scale reconnection layer and the relevant issues to be addressed in the future.
\end{abstract}

\maketitle

\section{\label{sec:intro}Introduction}
In space, solar, and astrophysical plasmas, magnetic reconnection is one of the primary mechanisms for converting magnetic energy into plasma kinetic energy and accelerating high-energy nonthermal particles~\cite{Zweibel2009Magnetic}. One remarkable example is solar flares, where magnetic reconnection is thought to trigger the release of a large amount of magnetic energy in the solar corona~\cite{Carmichael64,Sturrock66,Hirayama74,Kopp76} and drive the acceleration of a large number of nonthermal electrons and ions. These particles produce a broad range of nonthermal emissions (hard X-ray, microwave, and gamma-ray)~\cite{Lin1976Nonthermal,Chen2015Particle,Chen2018Magnetic,Chen2020NatAs,Gary2018Microwave,Vilmer2011Properties}. Some of the accelerated particles get released into interplanetary space as impulsive solar energetic particles~\cite{Reames1999,Wang2012Stat}. Understanding how particles are accelerated during reconnection and the resulting particle energy distributions have been central problems in the study of magnetic reconnection.

One ubiquitous feature revealed by observations is that the accelerated particles tend to develop power-law energy spectra~\cite{Miller1997Critical,Mason2002Spectral,Oka2018Electron}. Solar flare emissions often show power-law energy distributions and are presumably produced by nonthermal particles with power-law distributions~\cite{Sui2003Evidence,Krucker2010Measure,Oka2013Kappa,Oka2015Electron,Oka2018Electron,Gary2018Microwave}. Although other mechanisms may also accelerate particles, reconnection holds the promise to accelerate a large number of electrons~\cite{Drake2006Electron}, as suggested by some observations~\cite{Masuda1994Loop,Krucker2010Measure}. In addition, reconnection-accelerated particles may serve as an injection population to be further accelerated by other mechanisms, such as termination shocks~\cite{Guo2012Particle,Chen2015Particle,Kong2019}, plasma turbulence~\cite{Miller1993Ion,Miller1996Stochastic,Miller1997Critical,Petrosian2004Stochastic,Petrosian2012Stochastic} driven by reconnection outflow and large-scale Alfv\'en waves in the flare loops~\cite{Fletcher2008}. Thus, it is crucial to understand particle acceleration by solar flare reconnection. In-situ observations in space plasmas give more direct evidence of power-law formation in the reconnection regions. In Earth's magnetotail, electrons develop power-law energy spectra near the reconnection diffusion region~\cite{Oieroset2002Evidence,Oka2016}, inside magnetic islands~\cite{ChenLJ2009,Retino2008,Huang2012Electron}, and in the flux pile-up region~\cite{Imada2007,Fu2013NatPh}. In the solar wind, local enhancements of energetic particles and the formation of power-law energy spectra have been associated with contracting and merging small-scale flux ropes.~\cite{Khabarova2015,Khabarova2016,Khabarova2017,Zhao2018Unusual,Zhao2019Particle,Adhikari2019Role}. Motivated by these observations, many studies have been dedicated to investigating particle acceleration associated with magnetic reconnection. While the analytical studies have made predictions consistent with observations~\cite{Zank2014Particle,LeRoux2015Kinetic,Zhao2018Unusual,Zhao2019Particle,Adhikari2019Role}, the whole process can be very nonlinear and numerical simulations still have to provide convincing evidence for power-law formation during reconnection. Test-particle simulations (either in full orbit or guiding-center motions) usually generate hard power-law energy spectra, much harder than that observed in space and solar plasmas~\cite{Onofri2006Stochastic,Gordovskyy2010ParticleA,Gordovskyy2010ParticleB,Zhou2015Electron,Zhou2016Electron,Xia2018}. Comparing to the kinetic simulations of relativistic reconnection (the Alfv\'en velocity approaches the speed of light) that consistently produce hard power-law energy spectra~\cite{Sironi2014Relativistic,Guo2014Formation,Guo2015Efficient,Guo2019Determining,Werner2014Extent}, simulations in the nonrelativistic regime relevant to space and solar plasmas show persistent difficulties in obtaining power-law spectra~\cite{Drake2010Magnetic,Drake2013Power,Dahlin2014Mechanisms,Li2015Nonthermally,Li2017Particle}.

In this review, we focus on critical issues for understanding energetic particle acceleration and power-law formation in reconnection. First, what are the dominant particle acceleration mechanisms during magnetic reconnection? Traditionally, it is thought that particles are accelerated only by the non-ideal electric field near the reconnection X-line~\cite{Speiser1965}, which cannot explain the large number of electrons generated during solar flares. Recent studies have made significant progress showing that Fermi-type mechanisms are important for particle acceleration during reconnection~\cite{Drake2006Electron,Guo2014Formation,Dahlin2014Mechanisms,Li2015Nonthermally}. While the acceleration mechanisms are similar in different regimes, power-law energy spectra are only commonly seen in the relativistic simulations, suggesting that the plasma parameters (e.g., Alfv\'en velocity, plasma $\beta$, or guide field--the magnetic field component perpendicular to the reconnection plane) might play a significant role in the power-law formation. The broad distributions of these parameters in space and solar plasmas could explain why the power-law indices range broadly from 2 to 9~\cite{Lin1976Nonthermal,Krucker2010Measure,Oka2013Kappa,Oka2015Electron,Oka2018Electron,Effenberger2017Hard} (In the rest of this paper, a power-law energy spectrum is defined as $f(\varepsilon)=dN(\varepsilon)/d\varepsilon\sim\varepsilon^{-p}$, where $\varepsilon$ is particle kinetic energy, $dN(\varepsilon)$ is the number of particles in a spherical energy shell with thickness $d\varepsilon$ around $\varepsilon$, and $p$ is the spectral index). Second, how do particles propagate in the reconnection region? To gain more energy, particles have to be trapped in the acceleration region for a long time or access the acceleration region multiple times. Both depend on the waves and turbulence in the reconnection region and will become even more important in the 3D reconnection layer that can self-generate plasma turbulence~\cite{Bowers2007Spectral,Daughton2011Role,Liu2013Bifurcated,Dahlin2015Electron,Dahlin2017Role,Le2018Drift,Stanier2019Influence,Li2019Formation,Huang2016Turbulent,Beresnyak2017Three,Kowal2017Statistics}. This review summarizes some recent progress on these two issues and points out several problems to be addressed in the next step. In Section~\ref{sec:mechanisms}, we summarize particle acceleration mechanisms during reconnection and three closely linked views on the Fermi mechanism. In Section~\ref{sec:transport}, we summarize how 3D physics facilitate particle transport and acceleration. In Section~\ref{sec:power_law}, we present a model for explaining the power-law formation in 3D reconnection. In Section~\ref{sec:outlooks}, we point out the challenges in studying particle acceleration in a large-scale reconnection layer and the relevant issues to be addressed in the future.

\section{Particle Acceleration Mechanisms}
\label{sec:mechanisms}

\subsection{Particle Acceleration Sites}
\begin{figure}
    \includegraphics[width=\linewidth]{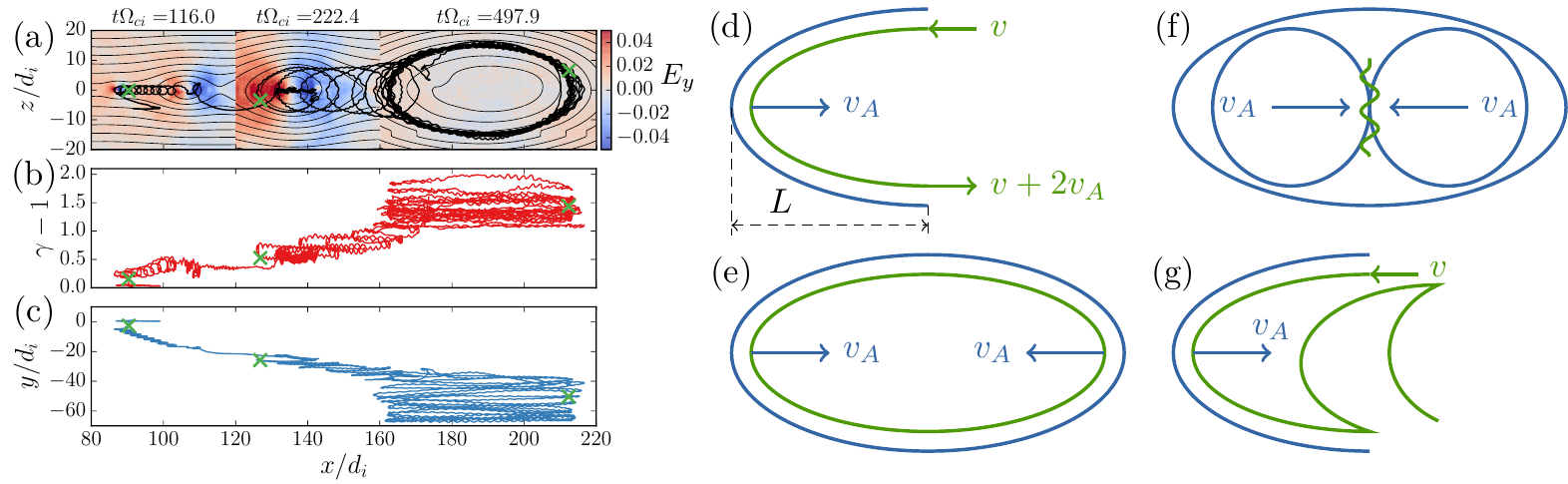}
    \caption{\label{fig:single_particle} Left: an electron trajectory showing the acceleration processes are associated with particle guiding-centering drift motions. (a) The trajectory in the simulation $x$--$z$ plane. The background is the out-of-plane electric field $E_y$ at three time slices. The green crosses indicate the positions of this electron at the three time slices. (b) The electron kinetic energy $\gamma-1$ (in the unit of $m_ec^2$) evolving with its $x$-position. (c) The electron's $x$-position versus its $y$-position. Reproduced with permission from Li et al., Astrophys. J, 843.1 (2017): 21. Copyright 2017 Institute of Physics (IOP). Right: Fermi mechanisms in reconnection. Blue curves indicate the magnetic field lines. Green curves indicate particle trajectories. $v$ is the particle velocity when entering the acceleration region. $v_A$ is the Alfv\'enic outflow velocity. (d) A single bounce off the reconnection outflow. Multiple bounces in (e) a contracting magnetic island, (f) island merging regions, and (g) a single reconnection exhaust.}
\end{figure}

The reconnection region is a natural particle acceleration site. Continuous magnetic flux comes into the reconnection region and deposits its free energy. At the same time, fresh, low energy particles are brought in through the reconnection inflow, get accelerated in the reconnection region, and escape as the outflow expels them out of the reconnection region. It has been demonstrated that the reconnection proceeds at a speed fast enough~\cite{Birn2001Geospace,Liu2017Why} to sustain the efficient magnetic energy conversion and particle acceleration observed in Earth's magnetotail and solar flares. During reconnection, the non-ideal electric field near the reconnection X-point can accelerate a small fraction of particles (mostly electrons)~\cite{Hoshino2001Suprathermal,Drake2005Production,Fu2006Process,Egedal2012Large,Egedal2015Double,Wang2016Mechanisms}. More importantly, reconnection-driven Alfv\'enic outflow on a much larger scale will induce motional electric fields that are more broadly distributed than the non-ideal electric field and can accelerate more particles~\cite{Guo2019Determining}. Recent studies have demonstrated that an elongated current sheet will break into a chain of magnetic islands (plasmoids) due to tearing instability~\cite{Shibata2001Plasmoid,Loureiro2007Instability,Bhattacharjee2009Fast,Huang2010Scaling,Comisso2016} or collapse of the reconnection exhaust~\cite{Liu2020Critical}. These magnetic islands tend to contract due to magnetic tension force, and they merge to form larger islands. Meanwhile, secondary current sheets are expected to form between the plasmoids, and new plasmoids can further develop in the new current sheets. This can repeatedly happen that gives rise to a hierarchical, fractal-like plasmoid structure~\cite{Shibata2001Plasmoid,Daughton2009Transition,Ji2011Phase,Daughton2012Emerging}. This fractal structure enables particles to gain energy at multiple sites in a reconnection region, such as the contracting magnetic islands~\cite{Drake2006Electron,Zank2014Particle,LeRoux2015Kinetic} and island merging regions~\cite{Oka2010Electron,Liu2011Particle,Drake2013Power,Nalewajko2015Distribution}. These acceleration regions have been identified in kinetic simulations by tracking particle trajectories~\cite{Drake2006Electron,Oka2010Electron,Guo2014Formation,Li2017Particle}, which include detailed information (e.g., position, velocity, and electric and magnetic fields) to characterize the acceleration history. Fig.~\ref{fig:single_particle}(a) shows the trajectory of an electron being energized in three contracting magnetic islands in a 2D particle-in-cell (PIC) simulation~\cite{Li2017Particle}. By carefully analyzing many such trajectories, we could differentiate the roles of these different acceleration regions. The trajectories have revealed that the parallel electric field near the reconnection X-line is not enough to accelerate particles to high energies but may provide an injection mechanism for further acceleration~\cite{Ball2019,Guo2019Determining,Kilian2020Exp}. High-energy particles are mostly accelerated in contracting or merging islands, and the island-merging regions seem to be the most efficient in accelerating electrons to the highest energy~\cite{Oka2010Electron,Nalewajko2015Distribution,Du2018Plasma}. Because of the multiple phases of accelerations, particles can be accelerated from thermal energy to nonthermal energies by reconnection without requiring external injection mechanisms like a X-point acceleration~\citep{Guo2019Determining,Kilian2020Exp}. These findings show that particle trajectories can help illustrate the acceleration processes. However, since one can only carefully analyze a small number of them, the statistical information on particle acceleration may be buried in the billions of other particles.

\subsection{\label{sec:fermi}Fermi mechanism in reconnection}

Although the reconnection region is complex with multiple acceleration regions and involves kinetic physics, the underlying particle acceleration mechanism turns out to be quite general. The Fermi mechanism has been proposed and demonstrated to be the dominant process for high-energy particle acceleration during reconnection~\cite{Drake2006Electron,Dahlin2014Mechanisms,Li2015Nonthermally}. Particles could gain energy through the Fermi mechanism when bouncing off the Alfv\'enic reconnection outflow driven by contracting magnetic field lines (Fig.\ref{fig:single_particle}(d)). The net energy gain due to one bounce is $\Delta\varepsilon=2m(vv_A + v_A^2)$, where $v$ is the initial particle speed, and $v_A$ is the Alfv\'en speed. Assuming the interaction takes about $\Delta t = 2L/v$ ($L$ is the characteristic length scale, e.g., the typical island diameter), the acceleration rate $\alpha\equiv\Delta\varepsilon/\varepsilon\Delta t\approx 2v_A/L$ if $v\gg v_A$, which is independent of particle energy, as predicted by the Fermi mechanism. In the non-relativistic regime, one bounce is usually not enough to accelerate particles to high energies, especially for electrons (which usually have $v\gg v_A$). To gain more energy, particles need to bounce multiple times off the flows. The multi-scale structures in a reconnection layer enable this scenario throughout the reconnection layer. Particles trapped in magnetic islands will continuously gain energy through bouncing off the flow driven by the contracting magnetic field lines~\cite{Drake2006Electron} (Fig.\ref{fig:single_particle}(e)) and merging magnetic islands~\cite{Oka2010Electron,Drake2013Power} (Fig.\ref{fig:single_particle}(f)). Recent findings have also shown that particles could interact with a single reconnection outflow multiple times to gain energies when trapped near the reconnection X-line by a parallel electric potential or the magnetic bottle~\cite{Egedal2015Double} (Fig.\ref{fig:single_particle}(g)). Alternatively, particles may experience pitch-angle scattering by self-generated turbulence or fluctuations in reconnection and bounce multiple times within a single reconnection exhaust~\cite{Dahlin2015Electron,Li2019Formation}.
\begin{figure}
    \includegraphics[width=\linewidth]{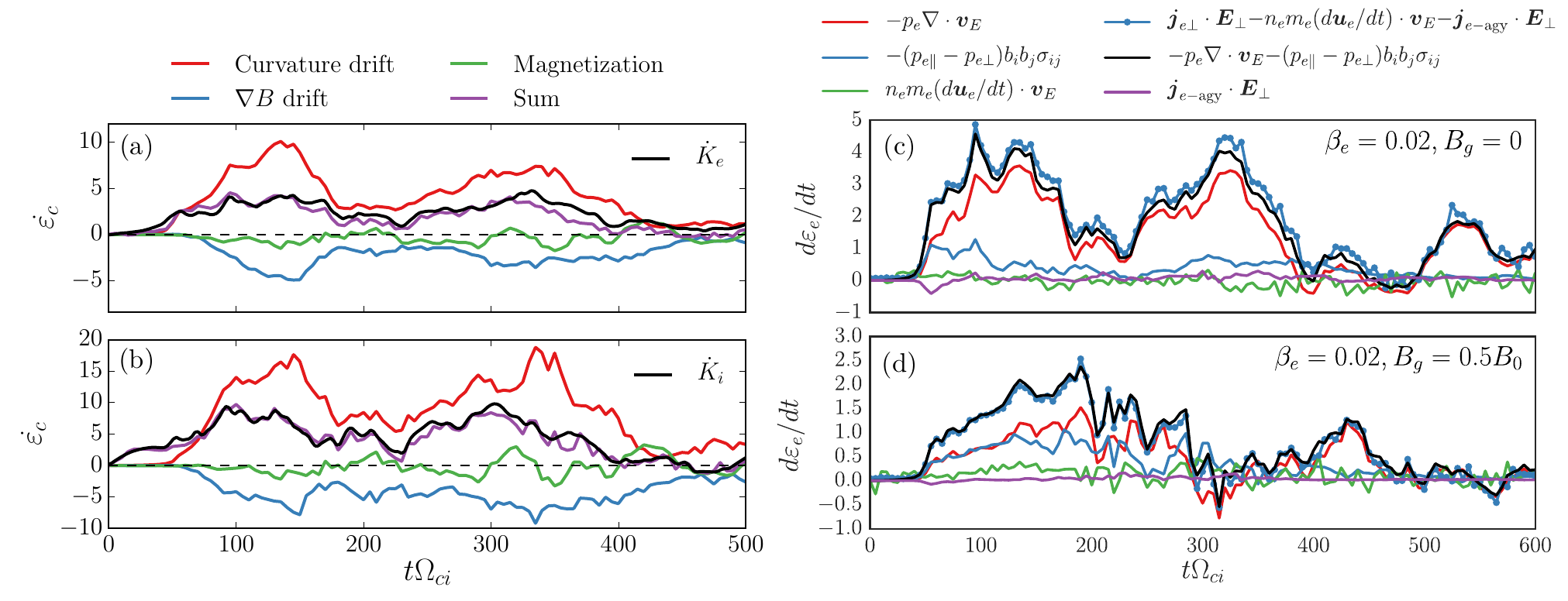}
    \caption{\label{fig:fluid_ene} Time evolution of different fluid energization terms over the entire simulation box. Left: the energization associated with guiding-center drift motions for (a) electrons and (b) ions in a simulation with $B_g=0$. Reproduced with permission from Li et al., Astrophys. J, 843.1 (2017): 21. Copyright 2017 Institute of Physics (IOP). Besides the listed three terms, the sum also includes the energization associated with polarization drift. $K_e$ and $K_i$ are particle kinetic energies. Right: the energization associated with flow compression and shear. Reproduced with permission from Li et al., Astrophys. J, 855.2 (2018): 80. Copyright 2018 Institute of Physics (IOP). $\vect{j}_{e-\text{agy}}\cdot\vect{v}_E$ is the energization associated with agyrotropic pressure tensor. (c) A simulation without a guide field. (d) A simulation with $B_g/B_0=0.5$.}
\end{figure}

When particles are well magnetized, the acceleration processes during reconnection can be well quantified by three closely linked models~\cite{Zank2014Particle,LeRoux2015Kinetic}. The first one is based on the conservation of the first and second adiabatic invariants, i.e., magnetic moment $\mu=mv_\perp^2/2B$ and parallel action $m\int v_\parallel dl$, where $v_\perp$ and $v_\parallel$ are velocities perpendicular and parallel to the local magnetic field, respectively. When magnetic islands contract or merge, the magnetic field lines tend to shorten~\cite{Drake2006Electron,Zank2014Particle}, increasing $v_\parallel$. Depending on whether the island area is conserved when islands contract, the magnetic field could get stronger or weaker. If the island area is conserved (circularization or incompressible case), the magnetic field and $v_\perp$ will decrease~\cite{Drake2006Electron,Drake2013Power,LeRoux2018Self,LeRoux2019}. If the island area shrinks (compressible case), the increasing magnetic field will lead to an increase of $v_\perp$. In contrast, the magnetic field and $v_\perp$ tend to decrease when two islands merge~\cite{Drake2013Power,Zank2014Particle}. Analytical models based on the conservation of these two adiabatic invariants have shown the formation of power-law energy spectra in a ``sea'' of magnetic islands~\cite{Bian2013Stochastic,Drake2013Power,Zank2014Particle,Zank2015Diffusive}.

The second model is based on particle guiding-center motions. Particle trajectories have shown that particles tend to drift in the out-of-plane direction (e.g., Fig.~\ref{fig:single_particle} (c)) when they gain energy (e.g., Fig.~\ref{fig:single_particle} (b)), suggesting that particle acceleration is associated with particle guiding-center drift motions, which include the parallel guiding-center velocity, $\vect{E}\times\vect{B}$ drift $\vect{v}_E$, grad-$B$ drift, inertial drift, parallel drift, and polarization drift~\citep{Northrop1963Adabatic, LeRoux2009Time, Webb2009Drift, LeRoux2015Kinetic}
\begin{align}
  \left<\vect{v}_g\right>_\phi = v_\parallel\vect{b} + \vect{v}_E +
  \frac{M}{q_s}\frac{\vect{B}\times\nabla B}{B^2} +
  \frac{m_sv_\parallel}{q_sB}\vect{b}\times\frac{d\vect{b}}{dt} +
  \frac{M}{q_s}(\nabla\times\vect{b})_\parallel +
  \frac{m_s}{q_sB}\vect{b}\times\frac{d\vect{v}_E}{dt},
\end{align}
where $\left<\dots\right>_\phi$ indicates gyrophase-average, $d/dt\equiv\partial/\partial t+(v_\parallel\vect{b}+\vect{v}_E)\cdot\nabla$, and $M\equiv m_s(\vect{v}_\perp - \vect{v_E})^2/2B$ is particle magnetic moment in the plasma frame where $\vect{v}_E=0$. The inertial drift (the fourth term on the right) includes the curvature drift $(m_sv_\parallel^2/q_sB)\vect{b}\times\vect{\kappa}$, where $\vect{\kappa}=\vect{b}\cdot\nabla\vect{b}$ is the magnetic curvature. In the guiding-center approximation, a single particle gains energy at a rate
\begin{align}
    \left<\dot{\varepsilon}\right>_\phi = q_s\vect{E}\cdot\left<\vect{v}_g\right>_\phi + M\frac{\partial B}{\partial t},
\end{align}
where the second term on the right is due to the conservation of the magnetic moment (betatron acceleration) and tends to be negative as the magnetic energy is released ($\partial B/\partial t<0$) in reconnection. Among the acceleration processes, the term associated with particle curvature drift is proportional to the particle parallel kinetic energy and is the Fermi mechanism in reconnection~\cite{Drake2006Electron}. Since the curvature drift is proportional to $m_sv_\parallel^2$ ($\approx m_sv^2/3$ for an isotropic particle distribution), the resulting acceleration rate  $\left<\dot{\varepsilon}\right>_\phi/\varepsilon\sim\vect{v}_E\cdot\vect{\kappa}$ is independent of particle energy. To compare the energization with kinetic simulations, we often integrate $\left<\dot{\varepsilon}\right>_\phi$ over electrons or ions to get the fluid energization terms~\cite{Parker1957Newtonian,Dahlin2014Mechanisms,Li2015Nonthermally,Li2017Particle,Wang2016Mechanisms}, obtaining $\vect{j}_s\cdot\vect{E}=\vect{j}_{s\parallel}\cdot\vect{E}_\parallel+\vect{j}_{s\perp}\cdot\vect{E}_\perp$, where $\vect{j}_s$ is total current density, of which the perpendicular component is (see Appendix B in~\citet{Li2019Particle} for derivation)
\begin{align}
    \vect{j}_{s\perp} = p_{s\parallel}\frac{\vect{B}\times(\vect{B}\cdot\nabla)\vect{B}}{B^4} +
    p_{s\perp}\frac{\vect{B}\times\nabla B}{B^3} -
    \left[\nabla\times\frac{p_{s\perp}\vect{B}}{B^2}\right]_\perp +
    \rho_s\vect{v}_E - n_sm_s\frac{d\vect{v}_s}{dt}\times
    \frac{\vect{B}}{B^2} \label{equ:jperp_drift},
\end{align}
where $p_{s\parallel}$ and $p_{s\perp}$ are parallel and perpendicular
pressures to the local magnetic field, respectively, $\rho_s$ is the charge density, $n_s$ is particle number density, $m_s$ is particle mass, $\vect{v}_s$ is the species flow velocity, and $d/dt\equiv\partial_t + \vect{v}_s\cdot\nabla$. The terms on the right show that the fluid energization is then associated with curvature drift, gradient drift, magnetization, and flow inertia. $\vect{v}_E$ does not contribute to the energization because $\vect{v}_E\cdot\vect{E}=0$. Fig.~\ref{fig:fluid_ene} left shows that the energization associated with curvature drift is dominant in the weak guide field limit. As the guide field $B_g$ increases, $\vect{j}_{s\perp}\cdot\vect{E}_\perp$ will be suppressed as the curvature of the magnetic field $|\vect{\kappa}|$ becomes smaller. Instead, the energization due to the parallel electric field $\vect{j}_{s\parallel}\cdot\vect{E}_\parallel$ will dominate when $B_g>B_0$~\cite{Dahlin2014Mechanisms,Li2017Particle}, where $B_0$ is the strength of the reconnection magnetic field.

The third model attempts to describe the energization in terms of flow compression and shear~\cite{LeRoux2015Kinetic,LeRoux2018Self,Li2018Roles,Du2018Plasma}. The former is the leading acceleration mechanism in energetic particle transport theory and diffusive shock acceleration~\cite{Parker1965Passage,Drury1983Introduction,Blandford1987Particle}. While the reconnection layer is assumed incompressible traditionally~\cite{Sweet1958,Parker1957}, recent MHD simulations have shown that compression effects are important in reconnection in low-$\beta$ and weak-$B_g$ regime~\cite{Birn2012Role,Li2018Large}. Kinetic models based on adiabatic invariants for reconnection acceleration have pointed out that compression could also be important for particle acceleration during reconnection~\cite{Zank2014Particle,LeRoux2015Kinetic,Montag2017Impact}.~\citet{Drury2012First} presented a reconnection acceleration model showing that the compression ratio between the outflow and inflow plasmas determines the power-law spectral index of reconnection-accelerated particles. These findings appear to be in contradiction with some previous theories that assume the reconnection layer is incompressible and compression acceleration is not in action~\cite{Drake2010Magnetic,Drake2013Power}. It turns out that the energization associated with guiding-center drift motions can be reorganized to
\begin{align}
    \vect{j}_{s\perp}\cdot\vect{E}_\perp & =
    \nabla\cdot(p_{s\perp} \vect{v}_E) - p_s\nabla\cdot\vect{v}_E
    - (p_{s\parallel}-p_{s\perp})b_ib_j\sigma_{ij} +
    n_sm_s\frac{d\vect{v}_s}{dt}\cdot{\vect{v}_E} \label{equ:comp_shear},
\end{align}
where $\sigma_{ij} = 0.5 (\partial_i v_{Ej} + \partial_j v_{Ei} -(2\nabla\cdot\vect{v}_E\delta_{ij})/3)$ is the shear tensor of $\vect{v}_E$ flow, and $p_s\equiv(p_{s\parallel} + 2p_{s\perp})/3$ is the effective scalar pressure. The second term on the right is due to flow compression and $\propto$ the scalar pressure. The third term on the right is due to flow shear and $\propto$ the pressure anisotropy. When the particle distribution is nearly isotropic, the flow shear is ineffective in energizing plasma but can provide a second-order acceleration term for high-energy particles~\cite{Earl1988Cosmic}, such as cosmic rays. Using kinetic simulations,~\citet{Li2018Roles} showed that the flow compression and shear effects capture the primary plasma energization (Fig.~\ref{fig:fluid_ene} right) and particle acceleration processes in reconnection. Compression energization dominates reconnection energization but becomes comparable with shear energization when there is a moderate guide field ($B_g/B_0\sim 0.5$). As $B_g$ increases, the plasma becomes less compressible because the guide field has to be also compressed when the plasma is being compressed. As a result, the compression energization becomes less efficient as $B_g$ increases. On the other hand, the pressure anisotropy tends to increases with the guide field~\cite{Le2013Regimes,Li2018Roles}. As a result, the shear energization (proportional to the pressure anisotropy) decreases slower with the guide field than compression energization. Overall, the joint energization due to flow compression and shear decreases with the guide field, providing an alternative explanation why the plasma energization is less efficient as $B_g$ increases. These results suggest that one may study particle acceleration in a large-scale solar flare reconnection site using the transport theory.

\section{\label{sec:transport}Particle spatial transport}

\subsection{Artificial Confinement in 2D Simulations}
\begin{figure}[htbp]
    \includegraphics[width=0.9\linewidth]{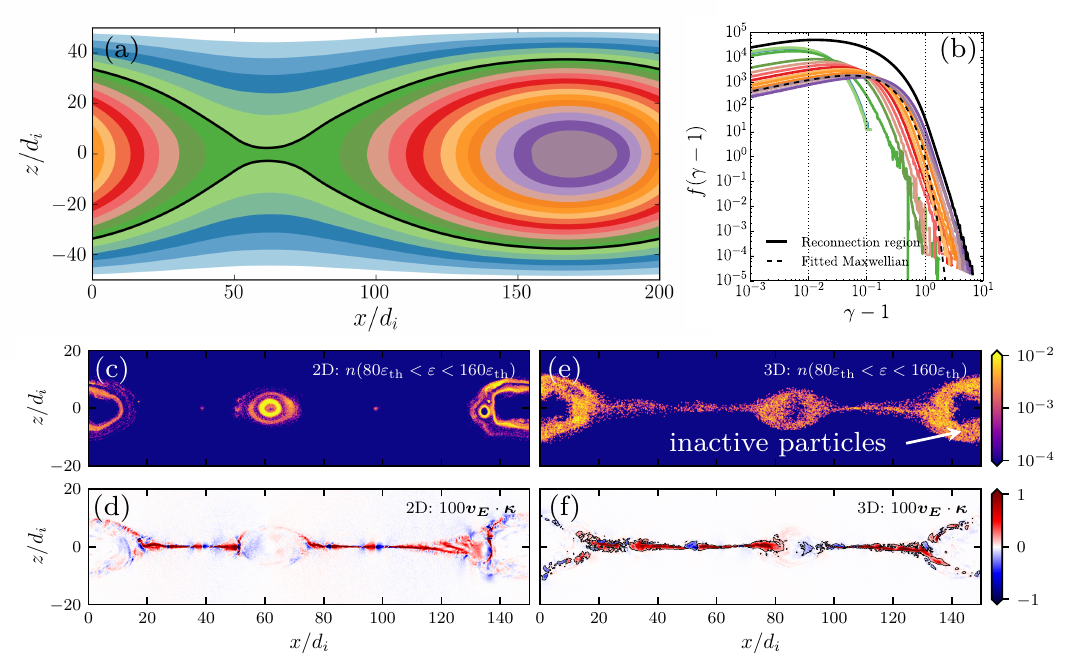}
    \caption{\label{fig:trapping_2d} Artificial particle trapping in 2D simulations. (a) The contour of the out-of-plane component of the vector potential $A_y$ in a run with $\beta_e=0.02$. The region between the black lines is the reconnection layer. (b) Electron energy spectra in these layers. The black dashed line shows the spectrum inside the reconnection layer bounded by the black lines in panel (a). The dashed line is the fitted Maxwellian for the spectrum in the innermost region. (a) \& (b): Reproduced with permission from Li et al., Astrophys. J, 843.1 (2017): 21. Copyright 2017 Institute of Physics (IOP). (c) The spatial distribution of high-energy electrons in the 2D simulation. (d) The acceleration rate associated with curvature drift $\vect{v}_{\vect{E}}\cdot\vect{\kappa}$ in a 2D simulation. Note that we multiply the rate by 100 just to make the color bar range from -1 to 1. (e) A slice of high-energy electron distribution in a 3D simulation. The arrow points out some electrons in the large flux rope that do not actively participate in the acceleration. (f) The corresponding acceleration rate associated with curvature drift in the 3D simulation. The black contour indicates the boundary of the major acceleration regions. (c)--(f): Reproduced with permission from Li et al., Astrophys. J, 884.2 (2019): 118. Copyright 2019 Institute of Physics (IOP).}
\end{figure}

Most kinetic simulations of reconnection were carried out in a 2D geometry due to computational constraints. In 2D reconnection, high-energy particles are mostly confined in magnetic islands due to the restricted particle motion across the magnetic field lines~\cite{Jokipii1993Perp,Jones1998Charged,Giacalone94}. Since particles are usually well magnetized and coupled with the field lines except in the diffusion regions, onion-like particle distributions form in the islands~\cite{Li2017Particle}. Particles entering the reconnection region earlier stay in an island's central region, while those  entering later will be in the outer region (Fig.\ref{fig:trapping_2d} (a)). Particles are adiabatically compressed and develop thermal-like energy distributions in each layer (Fig.~\ref{fig:trapping_2d} (b)). Those in the central region experience compression longer and thus have a higher temperature than that in the outer layers. Although some particles can still be accelerated to hundreds of times initial thermal energy, the particle distribution in the reconnection region is actually the superposition of these thermal-like distributions~\cite{Li2017Particle}. Furthermore, due to the confinement, high-energy particles cannot access the regions where the acceleration is most active (Fig.\ref{fig:trapping_2d} (c) \& (d)). The newly reconnected field lines in those regions are strongly bent and drive strong outflow ($\sim\vect{v}_E$) well aligned with the magnetic curvature $\vect{\kappa}$. Lower-energy particles accessing these regions will then have a higher acceleration rate associated with curvature drift $\vect{v}_E\cdot\vect{\kappa}$ than high-energy particles confined in the central region of magnetic islands. One may expect that merging islands will break the confinement and drive stronger particle acceleration and pitch-angle scattering through forming multiple X-lines. However, to access particles in a large island's central region, this island has to merge with one containing more magnetic flux, which will become less likely as fewer islands grow to large sizes.

\subsection{Fast Spatial Transport in 3D Reconnection}
\begin{figure}
    \includegraphics[width=\linewidth]{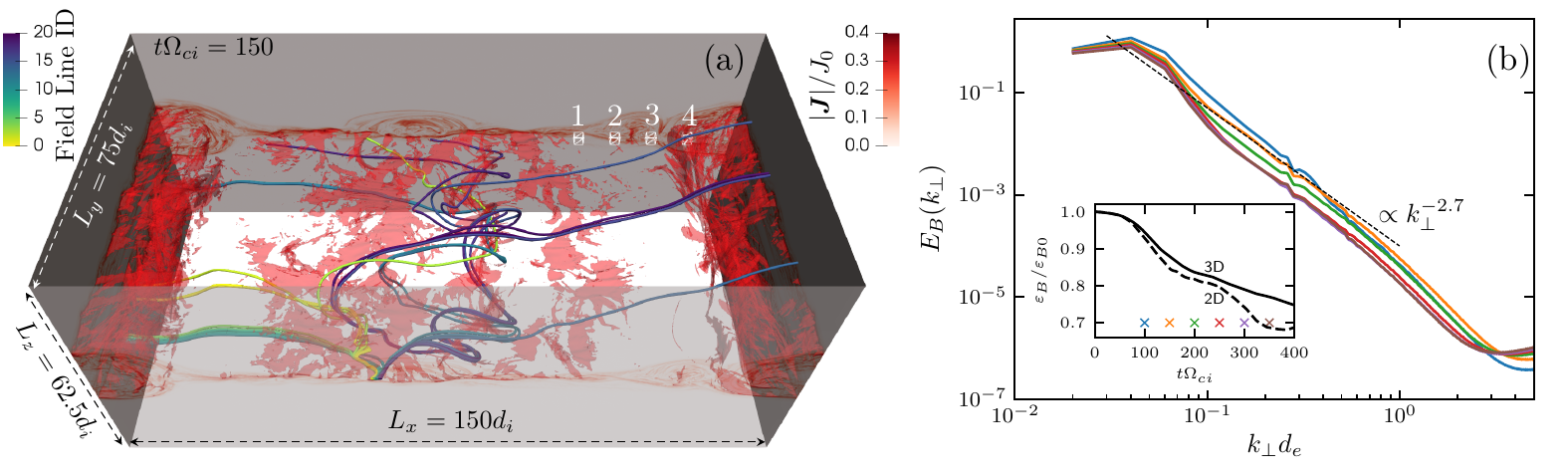}
    \caption{\label{fig:3dlayer} Turbulence and chaotic magnetic field lines in the 3D simulation. (a) 3D reconnection layer showing an isosurface of the current density with $|\boldsymbol{J}|/J_0=0.3$, and magnetic field lines starting from uniformly distributed points along a line of length $2d_i$. The field lines are color-coded with their seed identification numbers (IDs). The local electron energy spectra in the four whites cubes of $(2.3d_i)^3$ will be shown in Fig.~\ref{fig:spect_3d} (b). (b) Magnetic power spectra at five time slices indicated by the crosses in the embedded plot. The black dashed line indicates a power-law $\propto k_\perp^{-2.7}$. The inset also shows the time evolution of the magnetic energy $\varepsilon_B$ for both 2D and 3D simulations. $\varepsilon_{B0}$ is the initial magnetic energy. Reproduced with permission from Li et al., Astrophys. J, 884.2 (2019): 118. Copyright 2019 Institute of Physics (IOP).}
\end{figure}

Including 3D physics will help resolve these issues because magnetic flux ropes (the counterpart of magnetic islands in 3D) could be broken apart or even destroyed by plasma instabilities~\cite{Bowers2007Spectral,Daughton2011Role,Liu2011Particle,Huang2016Turbulent,Kowal2020}, such as oblique tearing and kink instabilities. As most of these instabilities require the third dimension, they are artificially suppressed in 2D simulations. The nonlinear growth and interaction of these instabilities will lead to a fragmented current layer filled with secondary flux ropes and current sheets (Fig.~\ref{fig:3dlayer}(a)). Magnetic field lines passing these regions will exponentially separate from each other and thus become chaotic~\cite{Daughton2014Computing,Dahlin2017Role,Le2018Drift,Stanier2019Influence,Li2019Formation,Yang2020,Guo2020Magnetic}. For example, the neighboring field lines in Fig.~\ref{fig:3dlayer}(a) quickly diverge from each other and access a broad region within just one crossing of the reconnection layer. The chaotic field lines enable different regions (e.g., flux ropes and reconnection exhausts) to be magnetically connected, allowing particles to access broader regions. The mixing of particles with different energies helps to develop truly nonthermal distributions locally (see next section for a discussion). Additionally, a 3D reconnection layer produces fluctuations that are likely to undergo both forward and inverse cascade~\cite{Bowers2007Spectral,Yang2020}, resulting in broadband turbulent fluctuations~\cite{Daughton2014Computing,Huang2016Turbulent,Kowal2017Statistics,Yang2020}. Fig.~\ref{fig:3dlayer}(b) shows that the magnetic power spectrum resembles a power-law distribution with a spectral index of about -2.7, consistent with that for reconnection-mediated turbulence in the collisionless regime~\cite{Loureiro2017Collisionless}. Other 3D simulations (either MHD or kinetic) have observed similar magnetic power spectra~\cite{Daughton2014Computing,Guo2015Efficient,Huang2016Turbulent} steeper than the classical Kolmogorov spectrum with a spectral index -5/3. If the energetic particles resonate with waves with wavelengths near the kinetic scales, such a steep spectrum will result in particle diffusion coefficients different from the quasi-linear theory assuming a Kolmogorov-like magnetic power spectrum~\cite{Jokipii1966,Giacalone1999Transport}.

\begin{figure}
    \includegraphics[width=\linewidth]{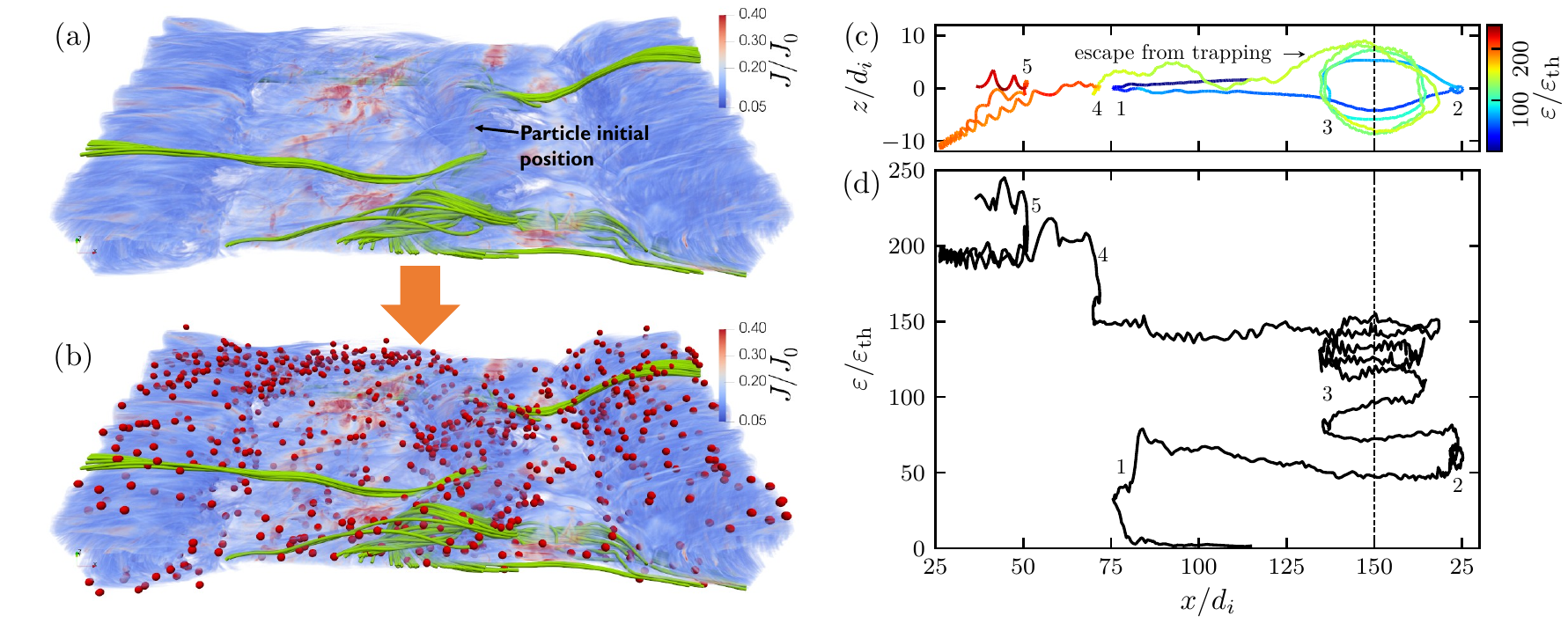}
    \caption{\label{fig:transport} Fast particle transport leads to stronger particle acceleration in a 3D reconnection layer. Panels (a) \& (b) are results from a test-particle simulation with the fields from the 3D PIC simulation as background. The particles are uniformly distributed in a ball with a radius of $d_e$ in the center of the flux rope (panel (a)). The magnetic field lines in green all pass through the ball. The red balls in (b) indicate their final positions. Panels (c) \& (d) show on electron tracer trajectory in the 3D PIC simulation. (c) The trajectory's projection on the $x$--$z$ plane. It is color-coded by its kinetic energy. The numbers 1--5 indicate five phases of acceleration. The arrow points out when the electron escapes from being trapped in the large flux rope. (d) $x$-position versus the electron kinetic energy. Reproduced with permission from Li et al., Astrophys. J, 884.2 (2019): 118. Copyright 2019 Institute of Physics (IOP).}
\end{figure}

High-energy particles in a 3D reconnection layer will resonate with the turbulent fluctuations and experience pitch-angle scattering~\cite{Jokipii1966}, preventing them from being confined in a flux rope for a long time. The snapshots of a test-particle simulation shown in Fig.~\ref{fig:transport} (a) \& (b) show how particles are quickly transported along the chaotic magnetic field lines in a turbulent reconnection layer. Instead of being trapped in the large flux rope, the electrons initially close to each other quickly fill the whole simulation domain after about two Alfv\'en crossing times.

The fast particle transport in 3D enables high-energy particles to access the regions where the magnetic energy is actively converted into plasma kinetic energy. Fig.~\ref{fig:trapping_2d}(e) shows the distribution of high-energy electrons in the 3D simulation shown in Fig.~\ref{fig:3dlayer}, and Fig.~\ref{fig:trapping_2d}(f) shows the corresponding acceleration rate associated with curvature drift acceleration. In contrast to the 2D results (Fig.~\ref{fig:trapping_2d}(c) \& (d)), high-energy electrons can access regions and have a large acceleration rate associated with curvature drift $\sim\vect{v}_E\cdot{\kappa}$. Thus, it is reasonable to expect that high-energy particle acceleration will become more efficient in 3D than in 2D. Fig.~\ref{fig:transport}(c) \& (d) show one electron trajectory traced in the 3D kinetic simulation. The electron gets three Fermi bounces (1--3) early in the simulation. It is then trapped in the large flux rope, and its acceleration gets slower and slower. In a 2D simulation, the electron would be confined in the large magnetic island until the island merges with other islands. However, in the 3D simulations, the electron can easily escape due to the chaotic magnetic field lines and self-generated turbulence. The trajectory shows it gets out of the flux ropes and experiences two more Fermi bounces in other regions (indicated by 4 \& 5 in Fig.~\ref{fig:transport}(c) \& (d)). These results show that the fast spatial transport due to chaotic field lines and self-generated turbulence helps particle acceleration in 3D simulations.

\section{The formation of power-law spectra}
\label{sec:power_law}

\subsection{Formation of Power-law Spectra in 3D}
\begin{figure}
    \includegraphics[width=\linewidth]{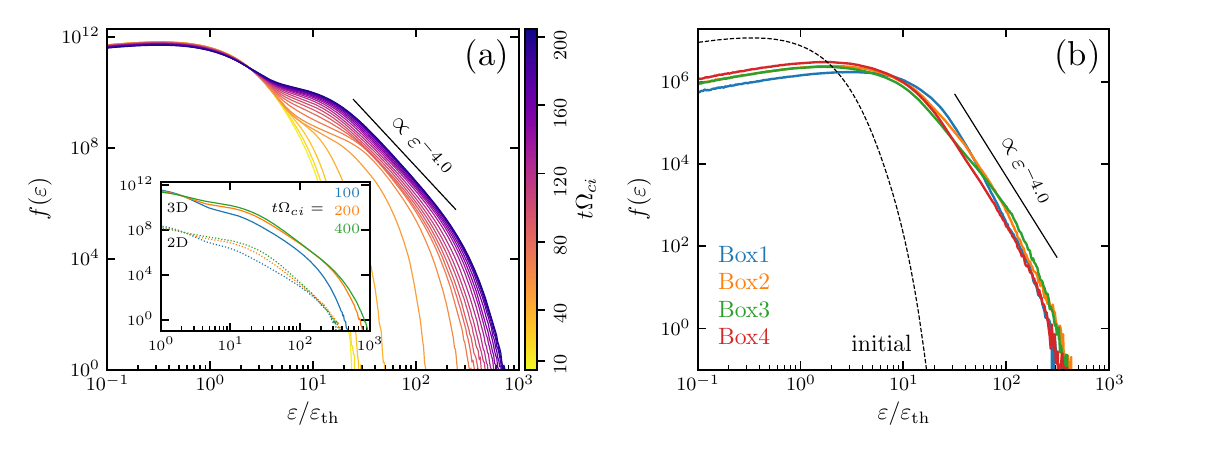}
    \caption{\label{fig:spect_3d} (a) Time evolution of the global electron energy spectrum in the 3D simulation with the inset comparing with the corresponding 2D simulation at three time slices. The black solid line indicates a power-law distribution with a spectral index $p=4$. We normalize $\varepsilon$ by the initial thermal energy $\varepsilon_\text{th}\approx0.015m_ec^2$. (b) Electron energy spectra for electrons in the four local boxes shown in Fig.\ref{fig:3dlayer} (a) at $t\Omega_{ci}=150$. Reproduced with permission from Li et al., Astrophys. J, 884.2 (2019): 118. Copyright 2019 Institute of Physics (IOP).}
\end{figure}

While plasma dynamics and reconnection rate have been the focus of most 3D simulations of magnetic reconnection~\cite{Daughton2011Role,Beresnyak2017Three,Huang2016Turbulent}, a few recent large-scale 3D kinetic simulations have been dedicated to studying particle acceleration in 3D reconnection~\cite{Guo2015Efficient,Guo2020Magnetic,Dahlin2015Electron,Dahlin2017Role,Li2019Formation}. In contrast to the relativistic simulations showing similar particle energy spectra in 2D and 3D reconnection~\cite{Sironi2014Relativistic,Guo2014Formation,Guo2015Efficient,Werner2017}, nonrelativistic simulations have shown stronger particle acceleration in 3D than in 2D because of the fast transport~\cite{Dahlin2015Electron,Dahlin2017Role,Li2019Formation}. Fig.~\ref{fig:spect_3d}(a) shows the time evolution of the electron energy spectrum in a 3D kinetic simulation~\cite{Li2019Formation}. The spectrum has a power-law tail $\propto\varepsilon^{-4}$, and it persists throughout the nonlinear reconnection phase until saturation. This result provides the first convincing evidence of forming the power-law energy spectrum in 3D nonrelativistic reconnection. Comparing with 2D reconnection, 3D reconnection accelerates electrons to higher energies. The high-energy tail keeps growing in the 3D simulation. In contrast, the maximum energy in the 2D simulation does not change much, and the flux keeps piling up at lower energies, causing the spectrum to become softer as the simulation evolves. This explains why the spectrum becomes very steep to the end of the 2D simulation shown in Fig.~\ref{fig:trapping_2d}(a) \& (b). The fast particle transport not only helps particle acceleration but also enhances particle mixing. Fig.~\ref{fig:spect_3d}(b) shows the local electron spectra at 4 different locations have similar power-law tails instead of different thermal-like distributions in 2D (Fig.~\ref{fig:trapping_2d}(b)).

\subsection{Power-law Formation Model}
Before looking into the simulation data, let us briefly review the formation of power-law spectra due to the Fermi mechanisms (see~\citet{Guo2020Recent} for a more detailed review). In the standard Fokker-Planck approach, the particle distribution evolves according to
\begin{equation}
  \partial_t f + \partial_\varepsilon(\alpha\varepsilon f) =
  \partial_\varepsilon^2 (D_{\varepsilon\varepsilon} f) - \frac{f}{\tau_\text{esc}} + \frac{f_\text{inj}}{\tau_\text{inj}},
  \label{equ:ene_cont2}
\end{equation}
where $D_{\varepsilon\varepsilon}$ is the energy diffusion coefficient, $\tau_\text{esc}$ is the escape time scale, $f_\text{inj}$ is the injected particle distribution, and $\tau_\text{inj}$ is the particle injection time scale. The acceleration rate $\alpha(\varepsilon)\equiv (d\varepsilon/dt)\varepsilon^{-1}=(\partial_t\varepsilon +\partial_\varepsilon D_{\varepsilon\varepsilon})\varepsilon^{-1}$ could describe first-order Fermi processes, direct acceleration in the reconnection diffusion region, and a first-order drift in energy $\partial_\varepsilon D_{\varepsilon\varepsilon}$ associated with second-order Fermi mechanisms. Since the first-order Fermi mechanism is thought to dominate particle acceleration processes in magnetic reconnection~\cite{Drake2006Electron}, we can ignore the energy diffusion $D_{\varepsilon\varepsilon}$ and the associated energy drift term ($\sim\partial_\varepsilon D_{\varepsilon\varepsilon}$) in $\alpha$. For particles with much higher energies than the injected particles, their energy distribution resembles a power-law with a spectral index~\cite{Drury1999}
\begin{equation}
\label{equ:pindex}
    p=1+\frac{1}{\alpha\tau_\text{esc}}+
    \frac{\partial\ln\alpha}{\partial\ln\varepsilon},
\end{equation}
which simplifies to $1+(\alpha\tau_\text{esc})^{-1}$ when both rates are
independent of particle energy~\cite{Guo2014Formation}. The spectral index is for particles in the acceleration region. For the energetic particles escaped from the acceleration region, the flux $f/\tau_\text{esc}$ has a power-law index
\begin{equation}
    p'=1+\frac{1}{\alpha\tau_\text{esc}}+
    \frac{\partial\ln\alpha}{\partial\ln\varepsilon} -
    \frac{\partial\ln\alpha_\text{esc}}{\partial\ln\varepsilon},
\end{equation}
where $\alpha_\text{esc}=\tau_\text{esc}^{-1}$ is the escape rate. Both $\alpha$ and $\alpha_\text{esc}$ (and thus $\tau_\text{esc}$) can be energy-dependent in general, but their energy-dependence should be the same for $p$ and $p'$ to be energy-independent. In reconnection, it is often assumed that $\tau_\text{esc}\sim L/v_A$ due to the advection loss by the reconnection outflow~\cite{Drake2013Power,Guo2014Formation,Arnold2021PRL}, where $L$ is the characteristic length scale (e.g., the typical island diameter). Since $\tau_\text{esc}$ does not depend on particle energy, $\alpha$ needs to be independent of particle energy as well. Therefore, to explain the formation of power-law spectra in nonrelativistic reconnection, one needs to demonstrate that the acceleration rate is indeed constant in the kinetic simulations. The model for explaining power-law formation in relativistic reconnection has shown that $\alpha\tau_\text{inj}$ should be $>1$ to obtain an extended power-law spectrum~\cite{Guo2014Formation,Guo2015Efficient,Guo2020Recent}, where $\tau_\text{inj}$ is particle injection time scale. This result indicates that simulations sustaining large $\alpha$ and long $\tau_\text{inj}$ will help power-law formation. Since the acceleration rate associated with the Fermi mechanism is $\sim\vect{v}_E\cdot\vect{\kappa}$, a larger Alfv\'en speed or magnetic curvature (or compression, see Section~\ref{sec:fermi}) will lead to stronger particle acceleration and the potential formation of power-law energy spectra in kinetic simulations. When the thermal speed is fixed, a larger Alfv\'en speed gives a lower plasma $\beta$. As the magnetic curvature decreases with the guide field, a larger magnetic curvature means a lower guide field. Thus, boosting $\alpha$ by performing simulations in low-$\beta$ and weak guide-field regimes will help obtain power-law spectra in a finite simulation time~\cite{Li2019Formation}.

\subsection{Acceleration and Escaped Rates in Reconnection}

\begin{figure}
    \includegraphics[width=\linewidth]{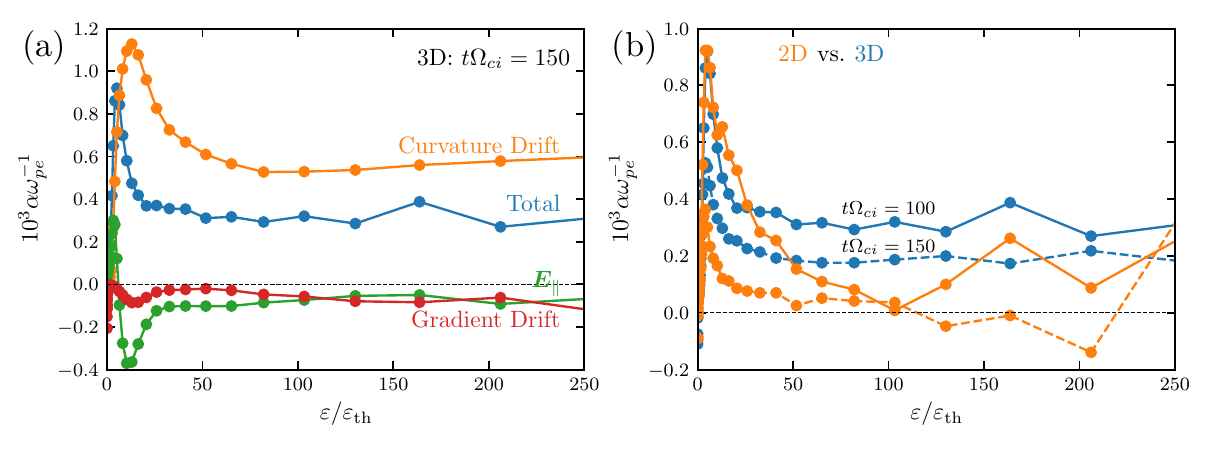}
    \caption{\label{fig:rates} Energy-dependent electron acceleration rate $\alpha$. (a) The contributions to $\alpha$ by the acceleration associated with curvature drift, gradient drift, and the parallel electric field in a 3D simulation. (b) $\alpha$ for the 2D (orange) and 3D (blue) simulations at two time slices. Reproduced with permission from Li et al., Astrophys. J, 884.2 (2019): 118. Copyright 2019 Institute of Physics (IOP).}
\end{figure}

Since kinetic simulations include self-consistent electric and magnetic fields at all particle positions, we can evaluate the energy-dependent particle acceleration rate $\alpha(\varepsilon)\equiv\left<\dot{\varepsilon}/\varepsilon\right>$, where $\left<\dots\right>$ is the ensemble average for particles in different energy bands, where $\dot{\varepsilon}=q\vect{v}\cdot\vect{E}$, and $\vect{v}$ is the particle velocity. We can also decompose the velocity into different guiding-center motions and calculate the acceleration rates associated with different acceleration mechanisms. Fig.~\ref{fig:rates}(a) shows the total and decomposed acceleration rates for electrons in a 3D simulation. The primary acceleration mechanism is a Fermi-type mechanism associated with particle curvature drift, the same as 2D results~\citep{Dahlin2014Mechanisms,Guo2014Formation,Li2015Nonthermally}. Both parallel electric field and grad-$B$ drift contribute much less than the curvature drift and even decelerate high-energy electrons. More importantly, since particles with different energies can access to the same acceleration regions due to the fast spatial transport and mixing, the total acceleration rate is nearly a constant in the 3D simulation. In contrast, due to the artificial confinement of high-energy electrons in 2D simulations, the electron acceleration rate sharply decreases with particle energy (Fig.~\ref{fig:rates}(b)) and even becomes negative for high-energy electrons later in the simulations. The energy dependence explains quantitatively why the electron energy spectrum keeps getting steeper in the 2D simulations (Fig.~\ref{fig:spect_3d} (a)).

\begin{figure}
    \includegraphics[width=0.6\linewidth]{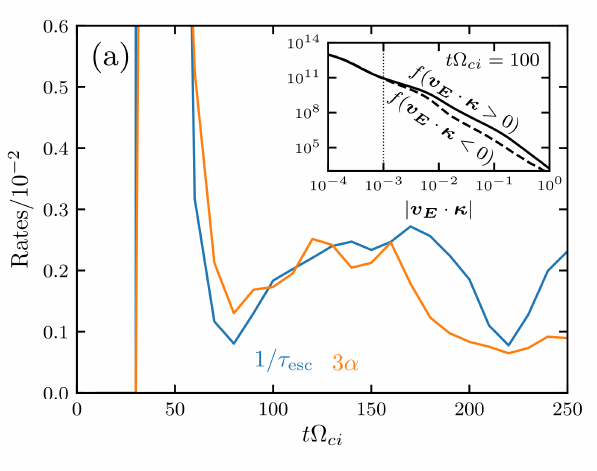}
    \caption{\label{fig:spect_model} An estimate of the power-law index by evaluating the acceleration rate $\alpha$ and the escape rate $1/\tau_\text{esc}$ for high-energy electrons ($\varepsilon>40\varepsilon_\text{th}$) in the major acceleration region, where $|\vect{v}_{\vect{E}}\cdot\vect{\kappa}|$ is larger than a threshold, as indicated in Figure~\ref{fig:trapping_2d} (f). (a) Time evolution of $3\alpha$ and $1/\tau_\text{esc}$. The embedded plot compares the distributions of the regions with negative and positive $\vect{v}_{\vect{E}}\cdot\vect{\kappa}$ at $t\Omega_{ci}=100$. The vertical dashed line indicates the chosen threshold 0.001 for $|\vect{v}_{\vect{E}}\cdot\vect{\kappa}|$. Regions with $|\vect{v}_{\vect{E}}\cdot\vect{\kappa}|<0.001$ do not contribute to the high-energy particle energization because positive $\vect{v}_{\vect{E}}\cdot\vect{\kappa}$ balance with negative values. Reproduced with permission from Li et al., Astrophys. J, 884.2 (2019): 118. Copyright 2019 Institute of Physics (IOP).}
\end{figure}

Particle escape processes are often not considered because most kinetic simulations have periodic or closed boundary conditions. A common assumption is that $\tau_\text{esc}\to\infty$ in these simulations because particles cannot leave the simulation box. However, if $\tau_\text{esc}\to\infty$, the power-law spectral index $p=1+(\alpha\tau_\text{esc})^{-1}$ will approach 1, which cannot explain the simulation results (Fig.~\ref{fig:spect_3d}). In reconnection, as the primary acceleration regions are only a fraction of the reconnection layer (Fig.~\ref{fig:trapping_2d}(d) \& (f)), effective escape is still possible when particles are out of these regions and do not actively participate in the acceleration processes (Fig.~\ref{fig:trapping_2d}(e)). Particles in large inactive magnetic structures (islands or flux ropes) are slowly accelerated and behave similarly to the escaped particles. Although some particles can be further accelerated when they are scattered out of these structures by turbulence~\citep{Dahlin2015Electron,Li2019Formation}, not all particles can be scattered and escape from the flux rope at the same time. Those escaped particles could be further accelerated when these structures merge, but the time scale between merging can be as long as the dynamical time scale. Thus, these large inactive magnetic structures serve as escape regions. By separating the acceleration and escape regions (Fig.~\ref{fig:trapping_2d}(f)), we can calculate the acceleration rate of the particles in the acceleration regions and their escape rate from the acceleration regions. The power-law spectrum is a dynamical balance between particle acceleration and escape, as predicted by Eq.~\ref{equ:pindex}. Fig.~\ref{fig:spect_model} shows that $\tau_\text{esc}^{-1}\approx3\alpha$ early in the simulation, and the resulted power-law index $p=1+(\alpha\tau_\text{esc})^{-1}\approx 4$ is similar to the simulation results. However, the rates deviate from each other after the reconnection outflows collide at the boundary at $t\Omega_{ci}\sim 150$. The acceleration rate sharply decreases after that, revealing why the power-law is short in the 3D simulation. Kinetic simulations with larger domains will likely sustain a more extended power-law spectrum, and simulations that can bypass the kinetic scales will also help the formation of extended power-law spectra~\cite{Li2018Large,Arnold2021PRL}.

\section{Outlooks}
\label{sec:outlooks}
\begin{figure}
    \includegraphics[width=0.6\linewidth]{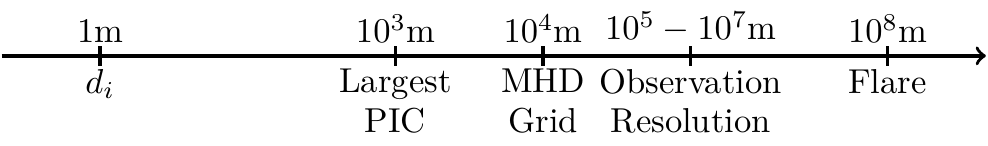}
    \caption{\label{fig:flare_scales} The large scale separation in solar flares. $d_i$ is the ion inertial length. The scale for the largest PIC is based on 2D simulations ($\sim 10^3 d_i$)~\cite{Dahlin2014Mechanisms,Li2015Nonthermally} and reduced plasma parameters (e.g., the mass ratio and light speed). Constrained by the computational cost, 3D simulations are even smaller than those 2D simulations~\cite{Li2019Formation}. The MHD grid scale is estimated based on some state-of-the-art MHD simulations of solar flares~\cite{Cheung2018Comprehensive,Dahlin2019,Shen2018}. The observation resolution depends on the emission band.}
\end{figure}

Recent kinetic simulations have made outstanding progress in understanding particle acceleration and power-law formation during magnetic reconnection. However, it is still incredibly challenging to address particle acceleration at global scales that are usually much larger than the kinetic scales. For example, Fig.~\ref{fig:flare_scales} shows that the flare scale is $\sim 5 $ orders of magnitude larger than kinetic simulations. The entire simulation domain of the largest PIC simulation is even smaller than the finest grid-scale of state-of-the-art MHD simulations, not to mention the observation scales. The large scale separation in typical space and solar plasmas (e.g., solar flares) prohibits using a single simulation to study particle acceleration and transport in a large-scale reconnection layer. To explain the observations, we need to incorporate the large-scale acceleration mechanisms learned from kinetic simulations (e.g., compression) into macroscopic energetic particle transport models to study particle acceleration and transport at observable scales~\cite{Li2018Large,Arnold2021PRL}. Here, we discuss several issues to address for tackling the global-scale acceleration problem in the future.

\subsection{Plasma Parameters and Simulation Setup}
Although kinetic simulations have shown evidence of the power-law formation in the nonrelativistic reconnection~\cite{Li2019Formation}, we have just started to explore the broad plasma parameter space (e.g., plasma $\beta$ and guide field). While 3D simulations in the low-$\beta$ and weak guide-field regimes have shown the power-law formation~\cite{Li2019Formation}, the results are only suitable for explaining power-law formation in solar flares and Earth's magnetotail. They are not proper for studying reconnection acceleration in the solar wind, where the plasma $\beta$ is considerably higher. Earlier studies have shown no evidence of substantial nonthermal particle acceleration in the reconnection exhausts observed in the solar wind~\cite{Gosling2005Absence}. However, recent studies have revived this topic and shown that reconnection-associated processes (e.g., contracting and merging magnetic flux ropes) can accelerate particles to develop nonthermal power-law energy spectra in the solar wind~\cite{Khabarova2017,Zhao2019Particle,Adhikari2019Role}. While analytical theories have shown that it is possible to develop power-law energy spectra in a ``sea'' of contracting and merging flux ropes even in high-$\beta$ plasmas~\cite{Drake2013Power,Zank2014Particle,LeRoux2015Kinetic}, numerical simulations are yet to demonstrate power-law energy spectra can indeed form in such plasmas. To test reconnection acceleration in the high-$\beta$ plasmas, performing simulations in these regimes and making $\tau_\text{inj}$ as large as possible is essential. Since $\tau_\text{inj}$ is limited by the reconnection duration and constrained by the boundary conditions (e.g., periodic boundaries), simulations with larger domains and open-boundary conditions will help. Considering the requirements of 3D simulations to achieve fast particle transport~\cite{Dahlin2015Electron,Dahlin2017Role,Li2019Formation}, studying power-law formation in high-$\beta$ regimes poses a challenge. For reconnection with a stronger guide field, it appears that the spectrum will become softer because the acceleration rate decreases with the guide field~\cite{Dahlin2014Mechanisms,Li2017Particle,Arnold2021PRL}. Whether this explanation is complete remains unclear due to the lack of studies on the escape mechanisms during reconnection and how the escape changes with different guide field.

Constrained by the computational cost, most kinetic simulations used reduced proton-to-electron mass ratio $m_i/m_e$ (e.g., 25) and ratio between the light speed and Alfv\'en speed. Such setup essentially reduces the scale-separation between electron and proton scales. While the reconnection rate is insensitive to the mass ratio~\cite{Li2019Particle}, particle acceleration mechanisms and power-law formation could be modified. For example, using a set of 2D simulations,~\citet{Li2019Particle} showed that a reduced mass ratio leads to a stronger electron acceleration by artificially enhancing the acceleration rate associated with curvature drift. If the mass-ratio dependence holds in 3D, the power-law formation processes and the power-law spectral index could deviate from previous results~\cite{Li2019Formation}. Therefore, we should be cautious when using the results from kinetic simulations to interpret the observations.

\subsection{Turbulence Properties and Particle Transport}
Despite the encouraging results from 3D kinetic simulations suggesting that fast particle transport is essential for the power-law formation in nonrelativistic reconnection~\cite{Li2019Formation}, no simulations have been dedicated to studying particle transport in the turbulent reconnection layer. To understand particle transport, it is important to quantify the turbulence properties relevant to particle acceleration first, including turbulence amplitude, power spectrum, anisotropy, compressibility, and correlation length~\cite{Zank2014Transport}. The turbulence amplitude $\delta B^2/B_0^2$ is the ratio of the fluctuating magnetic energy and the energy of the mean magnetic field. It represents how strong the magnetic field fluctuates and determines how strongly particles are scattered and whether second-order particle acceleration is significant~\cite{Petrosian2012Stochastic}. The turbulence amplitude might decrease with the guide field because the guide field is not dissipated in magnetic reconnection and because the magnetic energy conversion becomes less efficient as the plasma becomes less compressible when the guide field increases~\cite{Dahlin2016Parallel,Li2018Roles,Li2019Particle}. The other turbulence properties are subjects of intense debate~\citep{Huang2016Turbulent,Beresnyak2017Three,Boldyrev2017MHD,Kowal2017Statistics,Loureiro2017Role,Loureiro2017Collisionless,Mallet2017Disruption,Comisso2018MHD,Dong2018Role,Walker2018Influence}. In the MHD regime, there is no conclusion on whether these results are consistent with classical Goldreich–Sridhar theory for anisotropic turbulence in a homogeneous plasma permeated by a uniform magnetic field~\citep{GS95,Maron2001,Cho2000}. Compared with the MHD studies, studying turbulence properties in 3D kinetic simulations is even more challenging because the computation cannot cover a similar dynamical range as in MHD simulations due to the multi-scale nature. Although the turbulence spectrum is routinely shown in recent kinetic simulations~\citep{Daughton2014Computing,Guo2015Particle,Li2019Formation}, a good understanding of the anisotropy level, compressibility, and turbulence correlation length still needs more studies.

If a much larger PIC simulation is accessible, the particle diffusion coefficients can then be evaluated using these turbulence properties according to particle transport theories~\cite{Jokipii1966,Schlickeiser1998Quasi,Giacalone1999Transport,Chandran2000PRL,Matthaeus2003}. As a complement to this analysis, the diffusion coefficients can also be evaluated following the Taylor-Green-Kubo (TGK) forms using test particles in the 3D kinetic simulations. The comparison will tell us whether current transport theories can explain particle transport in reconnection-driven turbulence or new kinds of transport models are required~\cite{Zank2014Particle,Zank2015Diffusive,LeRoux2015Kinetic,LeRoux2018Self}. The results will guide the efforts in modeling particle acceleration and transport in large-scale reconnection layers, for example, by solving Parker's transport equation~\cite{Li2018Large}.

\subsection{Ion Acceleration}
Ion acceleration can be as efficient as electron acceleration during reconnection. For example, solar flare observations have shown well-correlated electron-generated hard X-ray flux and ion-generated gamma-ray flux~\cite{Shih2009RHESSI}, suggesting a similar acceleration mechanism for both electrons and ions. Also, heavy ions (especially Fe/O) often show substantial enhancement from their coronal abundance during solar flares~\cite{Mason2004}. Compared to electron acceleration, ion acceleration associated with magnetic reconnection is much less studied, especially in kinetic simulations. Although ions can reach a few thousands of times of the initial thermal energy~\cite{Li2017Particle,Li2019Particle}, the simulation domains are usually too small ($\sim100$ ion kinetic scales) for ions to have enough dynamical scale for developing power-law energy distributions. To study power-law formation for ions, hybrid kinetic simulations (kinetic ions and fluid electrons)~\cite{Stanier2015Role,Le2016} that do not resolve the electron kinetic scales are a better choice. The simulation box can be 5--10 times larger than the fully kinetic simulations, providing enough dynamical scales for ion acceleration.

To explain heavy-ion enhancement in flares, while previous theories mostly rely on wave-particle resonance in reconnection-driven turbulence~\cite{Miller1998,Petrosian2004Stochastic},~\citet{Drake2009Ion} proposed a novel mechanism by magnetic reconnection. When there is a finite guide field, the ions can be energized in a pickup process~\cite{Mobius1985} as they enter reconnection exhaust regions, and the heating of heavy ions (with low Q/M) is more efficient than protons. Although fully kinetic simulations have demonstrated the mechanisms to be effective~\cite{Drake2009Magnetic,Drake2014Onset,Knizhnik2011}, earlier works have been focused on ion heating instead of the acceleration of suprathermal ions and the formation of power-law spectra.

\subsection{Large-scale Acceleration Models}
Recently, significant efforts have been made in understanding particle acceleration in magnetic reconnection through macroscopic kinetic transport theories that incorporate the acceleration and transport processes learned from the fully kinetic simulations~\citep{Drury2012First,Drake2013Power,Zank2014Transport,LeRoux2015Kinetic,Li2018Large,Drake2019}. Inspired by these results,~\citet{Li2018Large} have studied compression particle acceleration in magnetic reconnection by solving the Parker (diffusion–advection) transport equation using velocity and magnetic fields from 2D MHD simulations of a low-$\beta$ reconnection layer. The compressible reconnection layer can give significant particle acceleration, leading to power-law particle energy distributions. These results show that one can use transport theories to study particle acceleration in a large-scale flare region. To improve the model, we need to incorporate more acceleration terms (e.g., flow shear and second-order acceleration) into the transport model and study whether they are essential for particle acceleration at large scales. In addition, the diffusion coefficients evaluated based on the turbulence properties will also help improve modeling particle transport in large-scale reconnection layers.

One potential limitation of the transport models (see~\citet{LeRoux2018Self} for an exception) is that energetic particles do not feedback to the background plasma. Solar flare observations indicate that magnetic reconnection can accelerate over 10\% of the entire electron population (more than $10^{36}$ electrons) in the flare region into nonthermal distributions~\cite{Oka2013Kappa,Oka2015Electron,Krucker2010Measure,Krucker2014Particle}. These energetic electrons thus contain a significant amount of energy that can affect the plasma dynamics. Consequently, the energetic particles' feedback to the plasma might be crucial for evolving the reconnection layer and associated particle acceleration. Recent results by the \texttt{kglobal} model~\cite{Drake2019,Arnold2019,Arnold2021PRL}, which evolves the guiding-center motions of electrons and includes the feedback to the MHD fluids, have shown the formation of power-law energy spectra in macroscale reconnection systems~\cite{Arnold2021PRL}. The power-law spectral indices are softer than the results in similar models but without feedback~\cite{Gordovskyy2010ParticleA,Gordovskyy2010ParticleB} and are similar to solar flare observations. Interestingly, these power-law spectra and their dependence on the guide field are similar to those obtained by solving Parker's transport equation~\cite{Li2018Large}. It suggests that the transport model, which does not feedback to the background plasmas but includes transport effects, is still very valuable for studying electron acceleration. To reveal the potential connection between these two approaches will require additional studies.

\section{Summary}
More and more evidence suggests that magnetic reconnection could be responsible for the nonthermal particle acceleration and power-law energy particle energy spectra observed in space and solar plasmas. Motivated by the observations, many studies (either analytical or numerical) have been focused on understanding how particles are accelerated in magnetic reconnection. We summarize the recent progress in particle acceleration mechanisms and spatial transport in the 3D reconnection and their roles in forming the power-law energy spectrum.

We reviewed how the multi-scale structure developed in a reconnection layer enables reconnection to be a natural site for accelerating particles from thermal to nonthermal energies. In 2D, the contracting magnetic islands and regions between merging magnetic islands are the most important for accelerating high-energy particles. The dominant particle acceleration mechanism has been proposed and demonstrated to be the Fermi mechanism using kinetic simulations. We summarized the three closely linked models for studying the Fermi mechanism. One is based on the two adiabatic invariants of magnetic moment and the parallel action and has become the foundation of many energetic particle models. The second one is based on individual particle's guiding-center drift motions and has revealed the energization associated with curvature drift as the dominant mechanism for high-energy particle acceleration. The third one is based on flow compression and shear and is equivalent to the other two models. It reveals the flow compression is critical for high-energy particle acceleration during reconnection and links reconnection acceleration with the classical energetic particle transport theories.

Although 2D studies have advanced our understanding of particle acceleration associated with reconnection, they are limited in studying nonthermal particle acceleration and power-law formation due to the artificial trapping of energetic particles. We reviewed recent progress on 3D physics that are relevant to particle acceleration and transport in reconnection. Both self-generated turbulence and chaotic magnetic field lines in a 3D reconnection layer can prevent particles from being trapped in local regions and enhance the spatial transport of energetic electrons. Consequently, energetic particles can access broader acceleration regions, gain more energy, and efficiently mix with other particles. The resulted particle energy distributions resemble power-laws globally and locally. We summarized a model for explaining the power-law formation in 3D based on the Fermi mechanism. The fast spatial transport enables particles with different energies to access the same acceleration regions, leading to the energy-independent acceleration rate, as often assumed for the Fermi mechanism. The power-law energy spectrum is due to the dynamical balance between particle acceleration and an effective escape caused by particles that do not actively participate in the acceleration.

We pointed out the challenges in addressing particle acceleration and transport in large-scale reconnection layers (e.g., solar flares) and several relevant issues to be addressed in the future to tackle the global-scale acceleration problem. First, how do the simulation results depend on the plasma parameters and simulation setup? We anticipate that both plasma $\beta$ and guide field will change the power-law formation processes. We should be cautious about using the simulation results based on reduced physics constants (e.g., proton-to-electron mass ratio and light speed) to interpret the observations. Second, what are the turbulence properties relevant to particle acceleration and transport in 3D reconnection? We anticipate that a good understanding of the turbulence amplitude, spectrum, anisotropy level, compressibility, and turbulence correlation length is critical in addressing this problem. Third, how are ions are accelerated, and what are the resulted ion energy distributions? We point out that the fully kinetic simulations might be too small to study ion acceleration, and hybrid kinetic simulations will be useful. One critical issue of studying ion acceleration is how the acceleration depends on the charge-to-mass ratio. Last but not least, what models do we have for studying particle acceleration in a large-scale reconnection layer? We point out recent progress in developing energetic particle transport models and including feedback from particles to plasma fluids for reconnection. Future studies should include more acceleration physics and transport effects in these models and perform cross-checking between different models.

\begin{acknowledgments}
    We thank the anonymous referees for very helpful and constructive reviews. We gratefully acknowledge discussions with Hui Li, Patrick Kilian, Qile Zhang, Joel Dahlin, Jim Drake, Adam Stanier, and William Daughton. Contribution from X.L. and Y.L. is based upon work funded by the National Science Foundation Grant No. PHY-1902867 through the NSF/DOE Partnership in Basic Plasma Science and Engineering and NASA MMS 80NSSC18K0289. F. G. acknowledges the support from Los Alamos National Laboratory through the LDRD/ER program and its Center for Space and Earth Science (CSES), and NASA programs through grant NNH17AE68I, 80HQTR20T0073, 80NSSC20K0627, and 80HQTR21T0005. Simulations were performed at the National Energy Research Scientific Computing Center (NERSC), at the Texas Advanced Computing Center (TACC) at The University of Texas at Austin, and with LANL institutional computing.
\end{acknowledgments}

\section*{DATA AVAILABILITY}
Data sharing is not applicable to this article as no new data were
created or analyzed in this study.




\bibliography{references}{}

\end{document}